\documentclass[apd,prd,reprint,showpacs,superscriptaddress,nofootinbib]{revtex4-1} 

\usepackage{amsmath} 
\usepackage{amssymb,latexsym} 
\usepackage{commath} 
\usepackage{gensymb} 
\usepackage{eucal} 
\usepackage{aas_macros}
\usepackage{algorithm}
\usepackage{algorithmic}
\usepackage{graphicx}
\usepackage{subfigure}
\usepackage{float}
\usepackage{enumerate}
\usepackage{hyperref}

\usepackage{color}

\renewcommand{\vec}[1]{\boldsymbol{#1}}

\newcommand{\mat}[1]{\boldsymbol{\mathbf{#1}}}
\newcommand{\RNum}[1]{\uppercase\expandafter{\romannumeral #1\relax}}

\newcommand{\K}{\ensuremath{\, {\rm K}}}
\newcommand{\mK}{\ensuremath{\, {\rm mK}}}

\newcommand{\MHz}{\ensuremath{\, {\rm MHz}}}

\begin{document}


\title{21~cm Signal Recovery via the Robust Principle Component Analysis}

\author{Shifan Zuo}
\affiliation{Key Laboratory of Computational Astrophysics, National Astronomical Observatories, Chinese Academy of Sciences, Beijing 100012, China}
\affiliation{School of Astronomy and Space Science, University of Chinese Academy of Sciences, Beijing 100049, China}
\email{sfzuo@bao.ac.cn}

\author{Xuelei Chen}
\affiliation{Key Laboratory of Computational Astrophysics, National Astronomical Observatories, Chinese Academy of Sciences, Beijing 100012, China}
\affiliation{School of Astronomy and Space Science, University of Chinese Academy of Sciences, Beijing 100049, China}
\affiliation{Center for High Energy Physics, Peking University, Beijing 100871, China}
\email{xuelei@cosmology.bao.ac.cn}

\author{Reza Ansari}
\affiliation{Universite ́ Paris-Sud, LAL, UMR 8607, F-91898 Orsay Cedex, France \& CNRS/IN2P3, F-91405 Orsay, France}
\email{ansari@lal.in2p3.fr}

\author{Youjun Lu}
\affiliation{Key Laboratory of Computational Astrophysics, National Astronomical Observatories, Chinese Academy of Sciences, Beijing 100012, China}
\affiliation{School of Astronomy and Space Science, University of Chinese Academy of Sciences, Beijing 100049, China}
\email{luyj@bao.ac.cn}

\date{\today}

\begin{abstract}
The redshifted 21~cm signal from neutral hydrogen (HI) is potentially a very powerful probe for cosmology, but 
a difficulty in its observation is that it is much weaker than foreground radiation from the Milky Way as well as 
extragalactic radio sources. The foreground radiation at different frequencies are however coherent along one line of 
sight, and various methods of foreground subtraction based on this property have been proposed. In this paper,    
 we present a new method based on the Robust Principal Component Analysis (RPCA) 
 to subtract foreground and extract 21~cm signal,  which explicitly uses both the low-rank property of the frequency covariance matrix (i.e. frequency coherence)  of the foreground and
 the sparsity of the frequency covariance matrix of the
  21~cm signal. The low-rank property of the foregrounds frequency covariance 
  has been exploited in many previous works on foreground subtraction, 
  but to our knowledge the sparsity of the frequency covariance of the 21~cm signal 
  is first explored here. By exploiting both properties in the RPCA method, in principle, 
  the foreground and signal may be separated without the signal loss problem. 
  Our method is applicable to both small patch of sky with the flat-sky approximation, 
  and to large area of sky  where the sphericity has to be considered. It is also easy to be 
  extended to deal with more complex conditions such as sky map with defects.
\end{abstract}

\maketitle

\section{Introduction}\label{S:intro}
Observation of the neutral hydrogen (HI) distribution through its
21~cm line radiation can provide very precious information on the history of
the Epoch of Reionization (EoR) \cite{2006PhR...433..181F,2010ARA&A..48..127M,2012RPPh...75h6901P, 2013ASSL..396...45Z} 
and the statistical properties of Large Scale Structure, 
which can be used to infer the nature of dark energy, dark matter, 
inflationary origin of the Universe \cite{2006astro.ph..6104P,2008PhRvL.100i1303C,
2008arXiv0807.3614A,2010ApJ...721..164S,2012A&A...540A.129A}. 
Detecting the 21~cm signal in a cosmological experiment is
a very challenging task, because there are various astrophysical foreground emissions, 
such as galactic synchrotron emission, free-free emission, recombination lines,
and the extragalactic radio sources including quasars, radio galaxies and galaxy clusters,
some of these are several orders of magnitude stronger than the 21 cm signal. 
For example, for the intensity mapping (IM) experiment where the 21 cm signal intensity of the large scale structure
is observed without resolving individual galaxies, the galactic synchrotron emission has 
$T_{b} \sim 10 \K$ while the HI signal $T_{b} \sim$ a few $\mK$ at $z \sim 1$. 
To separate the the HI signal from the foreground components 
one has to use their distinct statistical characteristics.  

The foreground spectra are expected to be
smooth in the frequency domain, while the redshifted 21~cm signal fluctuates randomly at different 
frequencies. Early works demonstrated that the 21~cm signal can be successfully 
extracted from such foregrounds by low order polynomial fitting in image space spectra
\cite{2006ApJ...650..529W,2008MNRAS.389.1319J,2009MNRAS.394.1575L},
or alternatively in Fourier space \cite{2009MNRAS.398..401L}, 
provided that the instrument response is smooth or well-known.
However, in the real world the instrument responses are not smooth, and are only known up to  the
 precision of calibration, which is itself a challenging task subject to many errors. More sophisticated 
 methods which can deal with the non-smooth apparent foregrounds in the observational data are therefore
 developed. In the so called blind or semi-blind methods, no specific parametric model for the foregrounds is 
 assumed, only generic features of the signal and foreground components such as their spectra
smoothness and frequency coherence are used. Examples 
include the low-order polynomial fitting; Principle Component
Analysis (PCA) or Singular Value Decomposition (SVD) \cite{2013ApJ...763L..20M,
2015MNRAS.454.3240B,2013MNRAS.434L..46S,2015ApJ...815...51S}
(to discriminate from the robust PCA method to be discussed below, we shall call it the {\it classic 
PCA} in the following);
Independent Component Analysis (ICA) \cite{2012MNRAS.423.2518C,2014MNRAS.441.3271W,
2015MNRAS.447..400A}; and the Generalized Morphological Component Analysis (GMCA) 
\cite{2013MNRAS.429..165C}, etc.
However, so far the redshifted 21~cm signal has not yet been positively detected in the various EoR 
and IM experiments, it is imperative to improve these methods further and 
explore  new approaches and methods . 

In recent years, many powerful techniques and efficient algorithms in signal processing 
come out from the compressed sensing (CS) researches  
\cite{Candes2006robust,Donoho06compressedsensing}, 
which offers a theoretical framework for simultaneously sensing and compressing finite dimensional
vectors by linear dimensionality reduction. It shows that sparse or compressible 
signals can be recovered from highly incomplete measurements by using appropriate
algorithms. The most prevalent structure used in CS is that
of {\it sparsity}. Sparsity implies that the signal
$x$ itself has only a small number of non-zero (or significantly
non-zero in the case of not exactly sparse but compressible case)
values, or $x$ can be sparsely represented in an appropriately chosen
basis or frame. Another often pursuit structure is the low-rankness of
matrices, where the rank is given by the number of non-zero singular
values of the matrix. Low-rank approximation has many important
applications for data compression, dimensionality reduction and so
on in areas such as computer vision, information retrieval, and
machine learning. 

An important role in these developments is played by convex
optimization \cite{Rockafellar1970,Boyd2004}, which is a subfield of
optimization, and has applications in a wide range of
disciplines. The convexity makes the optimization easier than the
general case since all its local minima must be global ones. Many
compressive sensing problems can be formulated into an optimization
problem, for example the most commonly encountered Basis Pursuit
(BP) problem (also called the $l_{1}$ minimization problem) in
compressive sensing
\begin{equation} \label{eq:BP}
  \min_{x} \, \norm{x}_{1} \qquad \text{subject to} \qquad y = Ax,
\end{equation}
can be solved by an equivalent convex optimization problem
\begin{equation} \label{eq:cv}
  \min_{x} \, \norm{y - Ax}_{2}^{2} + \lambda \norm{x}_{1},
\end{equation}
for some appropriate $\lambda >0$ (see the end of this section for notations).
Because of the universality of the compressive sensing techniques and
convex optimization methods, they found application  in many 
areas, including astronomy and astrophysics \cite{2008ISTSP...2..718B}.
Some CS techniques have been applied to radio astronomy, mostly in interferometer array image synthesis
\cite{2013ISPL...20..591C,2014MNRAS.439.3591C,EPFL-CONF-209227,2015A&A...575A..90G,2015JInst..10C8013G,2015arXiv150406847F,2016MNRAS.462.4314O,2016arXiv161008895O}. In fact,
even the classical H\"{o}gbom CLEAN algorithm \cite{1974A&AS...15..417H} and 
its multi-scale version \cite{1988A&A...200..312W} in radio image synthesis can be seen as matching pursuit
algorithms from the CS perspective \cite{1997A&AS..123..183L}.

In this work, we introduce the Robust Principle Component Analysis (RPCA) method,  which is a CS technique 
based on sparsity and low-rankness, to the problem of  21~cm signal and foregrounds separation. 
To our knowledge, this is the first conscious application of CS method in the 21~cm--foreground 
separation problem by taking advantage of the the sparsity and low-rank structure characteristics 
of the 21~cm signal and foregrounds frequency covariance information. It is also the first application of 
the RPCA method in 21cm foreground subtraction. While previous works has made use of the low-
rank property of the foreground, the present work first exploited the sparsity of the 21cm signal. 

The paper is organized as follows: We first describe the frequency
covariance structure of the foregrounds and the 21~cm signal, and how
the information of their distinct structures can be used to separate
them from each other in \autoref{S:fc} using the RPCA method. We introduce the RPCA
method and the algorithm for its solution in \autoref{S:alg}, and
show how this is done by a simulation in \autoref{S:sim}. Some details of the 21~cm signal and 
foreground simulation is given in Appendix \ref{S:gen}.
We then introduce the generalized Internal Linear
Combination (ILC) method to recovery the 21~cm signal from the
extracted 21~cm signal frequency covariance matrix in\autoref{S:21cm}, and compare the
performance of our introduced method to that of the classical PCA
method in \autoref{S:comp}. Finally we discuss the results
and its possible extensions in \autoref{S:dis}.

In this paper, we used various norms for vectors or
  matrices, for clarity, we summarize them here:
  \begin{description}
    \item[$\norm{\vec{x}}_{0}$] $l_{0}$-norm for vector or matrix $\vec{x}$, which
      is the number of non-zero elements of $\vec{x}$;
     \item[$\norm{\vec{x}}_{1}$] $l_{1}$-norm for vector $\vec{x}$,
      defined as $\norm{\vec{x}}_{1} \equiv \sum_{i} |x_{i}|$, if applied to a
      matrix the matrix is treated as a linear vector;
    \item[$\norm{\vec{x}}_{2}$] $l_{2}$-norm for vector $\vec{x}$,
      $\norm{\vec{x}}_{2} \equiv (\sum_{i} |x_{i}|^{2})^{1/2}$;
    \item[$\norm{\vec{x}}_{\infty}$] $l_{\infty}$-norm for vector
      $\vec{x}$, defined as $\norm{\vec{x}}_{\infty} \equiv \max |x_{i}|$, 
      when applied to a matrix the matrix is treated as a linear vector;
    \item[$\norm{\mat{A}}_{F}$] Frobenius norm for matrix $\mat{A}$, a generalization of the $l_2$ norm for matrix, 
    defined as  $\norm{A}_{F} \equiv (\sum_{ij} A_{ij}^{2})^{1/2}$;
    \item[$\norm{\mat{A}}_{*}$] Nuclear norm for matrix $\mat{A}$
      defined as the sum of its singular values. The nuclear norm
      can be interpreted as the $l_{1}$-norm of the vector of
      singular value of the matrix. 
  \end{description}

\section{The Robust PCA Method} \label{S:rpca}

\subsection{Frequency Correlations} \label{S:fc}
Consider a 21cm observation image data cube with two angular dimensions and one frequency 
dimension. The angular pixel indices are re-arranged as one index $p$, then the discrete multi-frequency 
sky maps is denoted as $x_{i}(p)$ at frequency $\nu_i$ and pixel $p$ as
\begin{equation} \label{eq:xip}
  x_{i}(p) = f_{i}(p) + s_{i}(p) + n_{i}(p),
\end{equation}
where $f_{i}(p)$ is the foreground, $s_{i}(p)$ the HI 21~cm
signal, and $n_{i}(p)$ the receiver noise contributions, or in vector form, 
\begin{equation} \label{eq:xp}
  \vec{x}(p) = \vec{f}(p) + \vec{s}(p) + \vec{n}(p).
\end{equation}

We assume that the 21~cm signal, astrophysical foreground and the
instrument noise are uncorrelated with each other,  and the noise in
different frequency channels are independent and can be modeled as a
zero-mean normal distribution, $n_{i}(p) \sim \mathcal{N}(0,
\sigma_{i}^{2})$, where $\sigma_{i}^{2}$ is the variance of the
noise in frequency channel $i$. Under this assumption, the $\nu -
\nu'$ covariance matrix of the noise will be a strictly diagonal
matrix $\mat{N} = \text{diag}\{\sigma_{1}^{2}, \cdots,
\sigma_{N_{\nu}}^{2}\}$. We have the $N_{\nu} \times N_{\nu}$ size
$\nu -\nu'$ covariance matrix of the observed data,
\begin{equation} \label{eq:cov}
  \mat{R} = \frac{1}{N_p} \langle \vec{x} \vec{x}^{T} \rangle = \mat{R}_{f} + \mat{R}_{\text{HI}} + \mat{N},
\end{equation}
where $\mat{R}_{f}$, $\mat{R}_{\text{HI}}$ and $\mat{N}$ are the $\nu - \nu'$
covariance matrix of $\vec{f}$, $\vec{s}$ and $\vec{n}$,
respectively, and the number of pixels $N_{p}$ is divided as an
normalization factor to make the value of the elements of the covariance matrix
roughly independent of pixelization. 
For a well constructed telescope system, to a first approximation, the frequency covariance matrix of 
the noise $\mat{N}$ may be assumed as diagonal, i.e. the noise could be described by 
a white noise model, and the different frequency bins are uncorrelated. Such noise  would be
indistinguishable from that of  the 21cm signal $\mat{R}_{\text{HI}}$,
 as both are sparse and their non-zero elements  concentrate on the diagonals. The noise
in this case can be suppressed by longer integration time or by cross-correlating with other 
tracer signal (e.g. the galaxy density obtained by optical observations). 
If we know the variance of the noise $\sigma_{i}^{2}$ in each frequency bin, we
could also subtract the frequency covariance matrix of the noise $\mat{N}$
from $\mat{R}$. In the following, to simplify the discussion, we shall not distinguish $\mat{N}$ and 
$\mat{R_{\rm HI}}$ any more, but treat them as an effective  $\mat{R_{\rm HI}}$,
and solve the foregrounds and 21~cm signal separation problem
\begin{equation} \label{eq:cov1}
  \mat{R} = \mat{R}_{f} + \mat{R}_{\text{HI}}.
\end{equation}

As radio point sources and diffuse foregrounds are all relatively smooth
along frequency and have a long frequency coherence, their $\nu -
\nu'$ covariance is expected to have very low ranks. A way to see
this is to note that $\mat{R}_{f}$ can be expressed as 
$$\mat{R}_{f} =\frac{1}{N_p} \mat{M}_{f} \mat{M}_{f}^{T},$$ 
where $\mat{M}_{f}$ is the $N_{\nu} \times N_{p}$ (foregrounds) sky data arranged as a matrix, and
$\mat{M}_{f}$ can be well modeled as a low-rank matrix. Its 
approximate SVD decomposition is
$$\mat{M}_{f} \cong \mat{U} \mat{\Sigma} \mat{V}^{T} =\sum_{i=1}^{r} \sigma_{i} \vec{u}_{i} \vec{v}_{i}^{*},$$
where $r \ll N_{\nu}$ is the rank of the matrix $\mat{M}_f$, and
$\sigma_{1} > \sigma_{2} > \cdots > \sigma_{r}$ are the non-zero singular 
values, and $\mat{U} = [\vec{u}_{1}, \cdots, \vec{u}_{r}]$,
$\mat{V} = [\vec{v}_{1}, \cdots, \vec{v}_{r}]$ are the corresponding
left and right singular vectors. This is the underlying principles
of the classic PCA/SVD foreground subtraction method
\cite{2013ApJ...763L..20M,2015MNRAS.447..400A,2015MNRAS.454.3240B,2013MNRAS.434L..46S,2015ApJ...815...51S}. 

On the other hand, the 21~cm signal has very short frequency coherence
because its frequency corresponds directly to redshift and thus
cosmic distance, its correlation diminishes
as the frequency difference $\Delta \nu$ increases. The correlation length $\Delta \nu$ for the 21~cm signal also
depends on the angular scale of observation, for the case of interest at $l \sim 100$,
this signal is uncorrelated beyond $\Delta \nu \sim 1$~MHz or even
less, while for $l \sim 10^{3}$ this occurs around $\sim 0.1$~MHz
\cite{2005MNRAS.356.1519B,2007MNRAS.378..119D}.  For the 21~cm signal, 
the $\nu - \nu'$ covariance matrix concentrates along the main diagonal, and the typical value of 
off-diagonal elements decay rapidly to essentially zero at a few MHz away. 
So the 21~cm $\nu - \nu'$ covariance matrix is a very sparse one, especially in broad band observations,
only elements along or near the diagonal have non-zero values.

\subsection{The Algorithm} \label{S:alg}
We see above that the $\nu - \nu'$ covariance matrix of the foregrounds and 21~cm signal 
have distinctly different characteristics, the 21~cm signal has a
very sparse structure, i.e., only elements along or near the the diagonal
are non-zero, while the $\nu - \nu'$ covariance matrix of foregrounds has low rank. 
Mathematically, separating such two components are exactly what the 
RPCA \cite{Candes:2011:RPC:1970392.1970395,NIPS2009_3704,Chandrasekaran2011}
is supposed to do, which tries to recover a low-rank component $L$ and a sparse
component $S$ from their superposition $M = L + S$ under some suitable
assumptions. In other words, the RPCA is to solve the problem
\begin{equation}
  \min \ \text{rank}(L) + \lambda \norm{S}_{0} \quad
  \text{s.t.} \quad  L + S = M, 
  \label{eq:rp0}
\end{equation}
over the (matrix) variables $L, S \in \mathbb{R}^{m \times n}$ 
and the regularization parameter $\lambda > 0$. This is a non-convex problem, since the minimization of
$\text{rank}(L)$ and $\norm{S}_{0}$ are non-convex and also NP-hard \cite{Natarajan:1995:SAS:207985.207987,Recht:2010:GMS:1958515.1958520}. Its convex relaxation is known as the Principal Component Pursuit (PCP), which is
\begin{equation}
  \min \ \norm{L}_{*} + \lambda \norm{S}_{1} \quad
  \text{s.t.} \quad  L + S = M, 
\label{eq:rp1}
\end{equation}
where the nuclear norm $\norm{L}_{*} = \sum_{i} \sigma_{i}(L)$ is
the sum of the singular values of the matrix, which is used as a
convex proxy for the non-convex $\text{rank}(L)$, and the
$l_{1}$-norm $\norm{S}_{1} = \sum_{ij}|S_{ij}|$ is used as a convex proxy for the
non-convex $l_{0}$-norm $\norm{S}_{0}$.
At first glance it may seem that for the PCP to work, one would
have to choose a right regularization parameter $\lambda$, but in
practice, the fixed choice $\lambda = 1 / \sqrt{\max{(m, n)}}$ works
very well for almost all cases \cite{Candes:2011:RPC:1970392.1970395}. 
In this sense, there is no tunable parameter in the PCP problem. 

Of course, in the most general case, there is an obvious degeneracy in the robust PCA problem
Eq.~(\ref{eq:rp0}) or its convex version -- the PCP problem
Eq.~(\ref{eq:rp1}), i.e., how to distinguish $L$ and $S$ if both are sparse and have low ranks. 
To separate the two, we make the following assumptions on $L$ and $S$:
\begin{enumerate}
\item $L$ is not sparse or ``spiky'' in the basis we start with. This is imposed 
  by requiring $L$ to be $\mu$-incoherent  \cite{Candes2008DBLP}:
  given the SVD of the low-rank component (with rank $r$) 
  $L = U\Sigma V^{*}$, let $U_{r}$ and $V_{r}$ denote the matrices consisting
  the first $r$ columns of $U$ and $V$ respectively, the
  incoherence parameter $\mu$ is a property of the matrix $L$, it is
  the smallest value that satisfies all the three inequalities \cite{Candes:2011:RPC:1970392.1970395}:
  \begin{eqnarray} \label{eq:muinc}
    \max_{i} \norm{U_{r}^{*} e_{i}}_{2}^{2} &\le& \frac{\mu r}{m}, \\
    \max_{i} \norm{V_{r}^{*} e_{i}}_{2}^{2} &\le& \frac{\mu r}{n}, \\
    \norm{U_{r} V_{r}^{*}}_{\infty} &\le& \sqrt{\frac{\mu r}{m n}}.
  \end{eqnarray}
  Satisfying the incoherence conditions means having a small value of
  $\mu$, this ensures that $L$ is
  not sparse \cite{Hornstein2011}.
\item The entries of $S$ are ``spread out'', i.e.
   for $\alpha \in [0,1)$, we assume $S \in \mathcal{S}_{\alpha}$, where $\mathcal{S}_{\alpha}$
 is defined as  $ \forall  i \in [n],  j \in [m]$,
  \begin{equation} \label{eq:Sa}
    \mathcal{S}_{\alpha} := \left\{ A \in \mathbb{R}^{m \times n} \
      \middle|  \ \|A_{(i, \cdot)}\|_{0} \leq \alpha n,  \|A_{(\cdot, j)}\|_{0} \leq \alpha m \right\}\nonumber\\
  \end{equation}
  In other words, $S$ contains at most a fraction $\alpha$ of non-zero
  entries per row and column. This guarantees that the probability of $S$ be
  low-rank is small. For example, if the non-zero elements of $A$ are
    mainly concentrate on the diagonal, as the case of the
    frequency covariance matrix of the 21~cm signal we are discussing
    it satisfies the condition $\mathcal{S}_{\alpha}$.
 \end{enumerate}
The frequency covariance of the foregrounds is low-rank
  but not sparse, satisfying condition 1; and the frequency covariance
  matrix of the 21~cm signal is sparse but not low-rank, satisfying
  condition 2. Under such assumptions, the PCP problems is solvable and convergence
guaranteed (c.f. Theorem 1.1 of
\cite{Candes:2011:RPC:1970392.1970395}).

The PCP problem is a convex optimization problem, many
off-the-shelf algorithms and tools are available for its solution 
\cite{LiuV09,Mohan:2012:IRA:2503308.2503351,Boyd2003,Lin2009,Eckstein2012,Boyd2011}. 
New algorithm customized for the RPCA and the PCP problem have also been developed, 
which are generally faster, more robust to corruptions or outliers, or more
scalable for large problems \cite{2010arXiv1009.5055L,NIPS2014_5430,Kang2015,YiPCC16}. 
Here we use an augmented Lagrange multiplier (ALM) algorithm \cite{2010arXiv1009.5055L,Yuan2009} to solve the
PCP problem Eq.(\ref{eq:rp1}), which is faster and more
accurate than, e.g., the Accelerated Proximal Gradient (APG) method \cite{Lin2009}. 
The ALM method operates on the augmented Lagrangian
\begin{eqnarray} \label{eq:lag}
  l(L, S, Y) &=& \norm{L}_{*} + \lambda \norm{S}_{1} + \langle Y, M -  L - S \rangle \nonumber\\
  &+& \frac{\mu}{2} \norm{M - L - S}_{F}^{2}, 
\end{eqnarray}
where $Y$ is an Lagrange multiplier matrix and $\mu$ is a positive
scalar. A generic Lagrange multiplier algorithm would solve the PCP
problem by repeatedly setting 
\begin{equation} \label{eq:sub-alm}
  (L_{k}, S_{k}) = \text{arg} \, \min_{L, S} \, l(L, S, Y_{k})
\end{equation}
and then updating the Lagrange multiplier matrix via 
$$Y_{k+1} = Y_{k} + \mu (M - L_{k} - S_{k}).$$
Two ALM methods had been proposed to solve the RPCA problem\cite{2010arXiv1009.5055L}: 
the exact ALM (EALM)  method has a pleasing Q-linear convergence speed, and a slight
improvement over the exact ALM leads an inexact ALM (IALM) method,
which converges practically as fast as the exact ALM, but the
required number of partial SVDs is significantly less. 
Here, we will use the IALM method, since the algorithm is
easy to implement, and performs excellently on a wide range of
problem settings without need of  tuning parameters. 

The IALM method does not solve Eq.~(\ref{eq:lag}) exactly, rather, it alternately 
updates $L_{k}$ and $S_{k}$ by solving a sequence of convex programs of 
$\text{min}_{L} \, l(L, S, Y)$ and $\text{min}_{S} \, l(L, S, Y)$ while keeping the other matrix variables
fixed. By doing so, both problems have very simple
and efficient closed solutions. Let $\mathcal{S}_\tau : \mathbb{R} \to
\mathbb{R}$ denote the shrinkage operator (also called the
soft-thresholding operator), 
$$\mathcal{S}_\tau[x] =
\text{sgn}(x) \, \text{max}( |x| - \tau, 0 ),$$ 
and extend it to matrices by applying it to each element. It is
easy to show that the solution to the first problem is
\begin{equation} \label{eq:minS}
  \text{arg} \, \min_{S} \, l(L, S, Y) = \mathcal{S}_{\lambda \mu^{-1}}(M
  - L + \mu^{-1} Y).
\end{equation}
Similarly, for matrix $X$, let $\mathcal{D}_\tau(X)$ denote the
singular value thresholding operator given by \cite{Cai2010a}
$$\mathcal{D}_\tau(X) =U \mathcal{S}_\tau(\Sigma) V^* ,$$ 
where $X = U \Sigma V^*$ is its singular value decomposition, it can be shown that
\begin{equation} \label{eq:minL}
  \text{arg} \, \min_{L} \, l(L, S, Y) = \mathcal{D}_{\mu^{-1}}(M
  - L + \mu^{-1} Y).
\end{equation}

The above method is summarized in Algorithm \autoref{alg:RPCA_ALM}, proof of
its convergence is given in  \cite{2010arXiv1009.5055L}.
The iteration of the algorithm can be stopped once 
\begin{equation}
\norm{M - L - S}_{F} \le \delta \norm{M}_{F},
\label{eq:ALMconverge}
\end{equation}
with a sufficiently small $\delta$, where the Frobenius norm  $\norm{M}_{F}^{2} = \sum_{ij} M_{ij}^{2}.$

\begin{algorithm}[H]
\caption{\bf (The PCP problem by ALM)}
\begin{algorithmic}[1]
\STATE {\bf initialize:} $S_0 = Y_0 = 0, \mu > 0, \delta > 0$.
\WHILE{$\norm{M - L - S}_{F} > \delta \norm{M}_{F}$ } 
\STATE \hspace{2 mm} compute $L_{k+1} = \mathcal{D}_\mu^{-1}(M - S_k + \mu^{-1} Y_k)$;
\STATE \hspace{2 mm}  compute $S_{k+1} = \mathcal{S}_{\lambda \mu^{-1}}(M - L_{k+1}
+ \mu^{-1} Y_k)$;
\STATE \hspace{2 mm} compute $Y_{k+1} = Y_k + \mu (M - L_{k+1} - S_{k+1})$;
\ENDWHILE
\STATE {\bf output:} $L, S$.
\end{algorithmic}
\label{alg:RPCA_ALM}
\end{algorithm}

In the problem of foregrounds and 21~cm signal component separation, as discussed above, 
the assumption that the low-rank component (the frequency covariance
matrix of the foregrounds) is not ``spiky'' and the entries of the
sparse one (the frequency covariance matrix of the 21~cm signal) are
``spread out'' is obviously satisfied.
In contrast to the usual application of the RPCA method and the PCP problem 
where recovery of the low rank matrix $L$ is the primary goal, here the sparse component $S$ 
is what we really want, as it will provide a good estimate for the HI 21~cm
frequency covariance matrix $\mat{\hat{R}}_{\text{HI}}$. The low
rank component $L$ may also be of interest as it gives a good
approximation of the foreground frequency covariance matrix
$\mat{\hat{R}}_{f}$, which as a byproduct, actually provides an 
estimate of the foreground from the observation itself. 
Note that the matrix $\mat{R}$ is symmetric positive definite in this problem, as it is the
frequency covariance matrix of the observed sky, but the RPCA
method and the PCP algorithm do not require this, the matrix $M$ in Eqs.(\ref{eq:rp0}) or (\ref{eq:rp1})
can be non-symmetric, so the same method can be applied in the case of cross-correlation with other observation, 
and even non-square matrix in the general case. 

\begin{figure*}[htbp]
  \centering
  \includegraphics[trim=180 10 180 10,clip,width=0.4\textwidth]{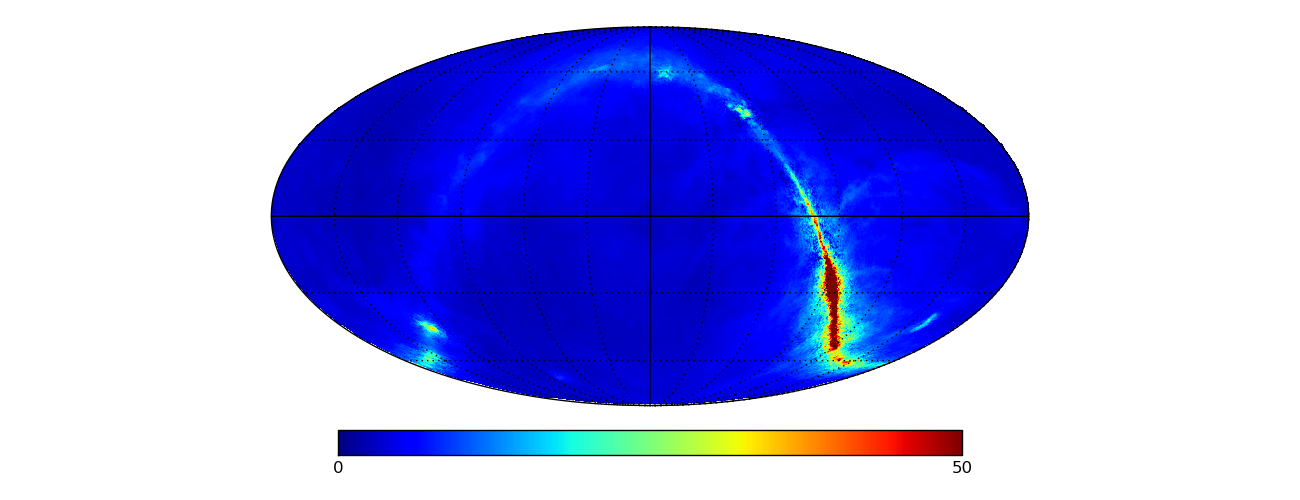}
  \includegraphics[trim=180 10 180 10,clip,width=0.4\textwidth]{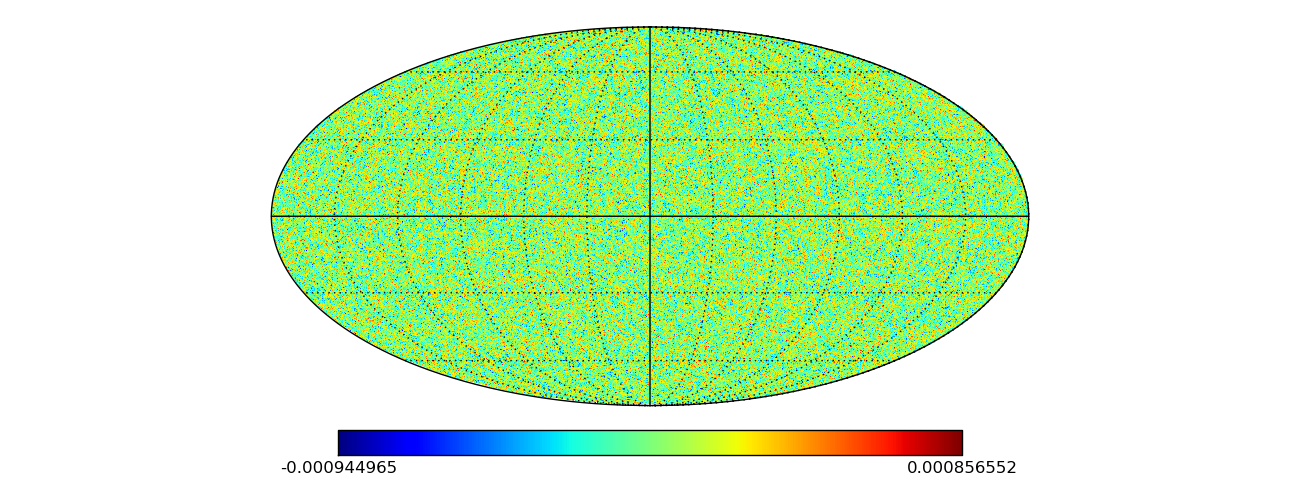} \\
  \includegraphics[trim=180 10 180 10,clip,width=0.4\textwidth]{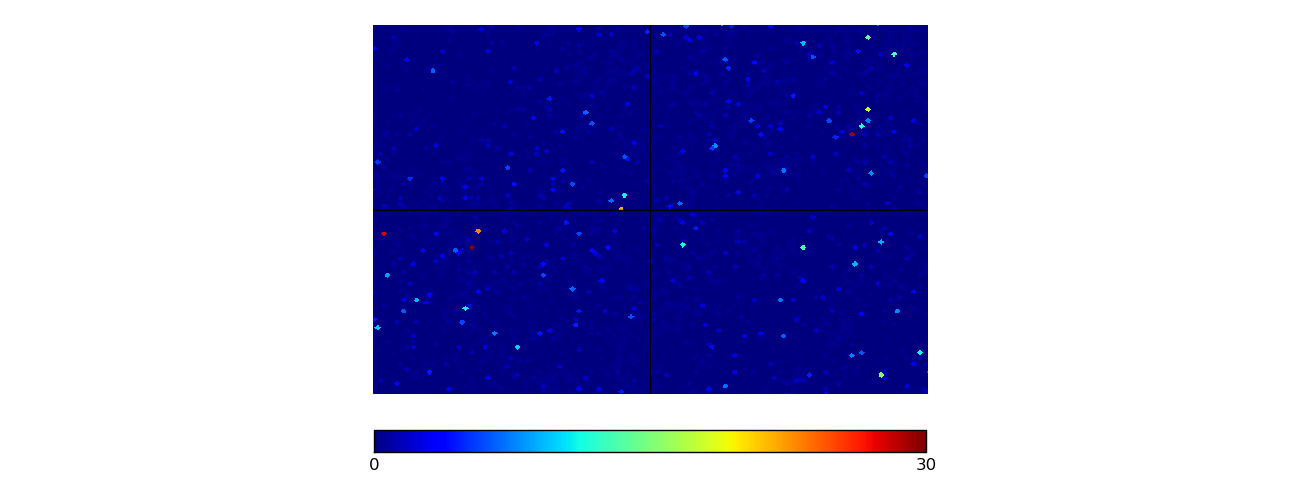}
  \includegraphics[trim=180 10 180 10,clip,width=0.4\textwidth]{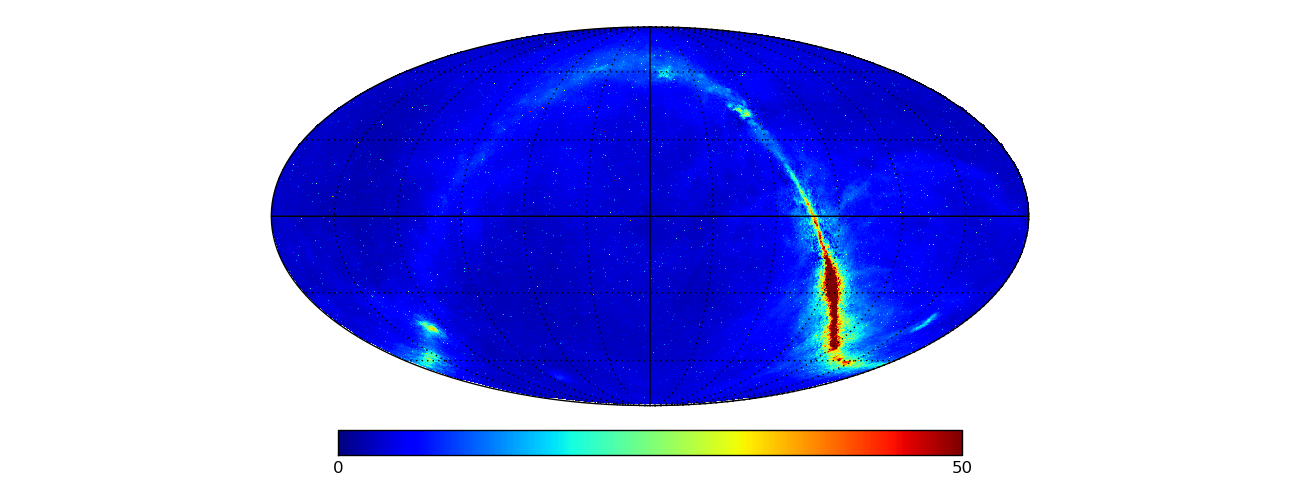}
  \caption{Input sky maps at 750 MHz. From left to right, top to bottom, the brightness temperature of the simulated Galactic synchrotron emission, 21~cm signal, extragalactic point sources and the sum of the three components, respectively. To better visualize the point source, in the 
 bottom left panel a small part of the map, ($-15\degree \le \alpha \le 15\degree$, 
 $-10\degree \le \delta \le 10\degree$), is shown .
  }
  \label{fig:maps}
\end{figure*}

For comparison, the frequently used classical PCA can be expressed as solving the (non-convex) problem
\begin{equation}
  \min \, \norm{M - L}_{F} \quad  \text{s.t.} \quad \text{rank}(L) \le k, 
  \label{eq:cp}
\end{equation}
with problem data $M$,  variable $L$, and integer $k \ge 1$. In the
classical PCA, $M$ is approximated as a low rank matrix $L$ with
the Frobenius norm error minimized. It can be efficiently solved with the SVD method, 
and  works well when the error is small and distributed as independent and identically 
Gaussian. But the classical PCA is not robust  when the data is grossly corrupted, i.e., with 
large outliers.  Also it is easy to fail in cases when a matrix is only a few sparse terms away from being low-rank,
especially if the sparse terms have large magnitudes. By contrast,
RPCA decomposes a matrix into the sum of a low-rank matrix and
a sparse matrix, thereby separates out the sparse errors, the
entries in the sparse error matrix $S$ can have arbitrary large
magnitude as long as its support is sufficiently small. Performing
this separation prevents the sparse errors from obscuring the
low-rank component \cite{Candes:2011:RPC:1970392.1970395}.

\subsection{Simulation} \label{S:sim}

\begin{figure*}[htbp]
  \centering
  \includegraphics[trim=30 10 30 10,clip,width=0.4\textwidth]{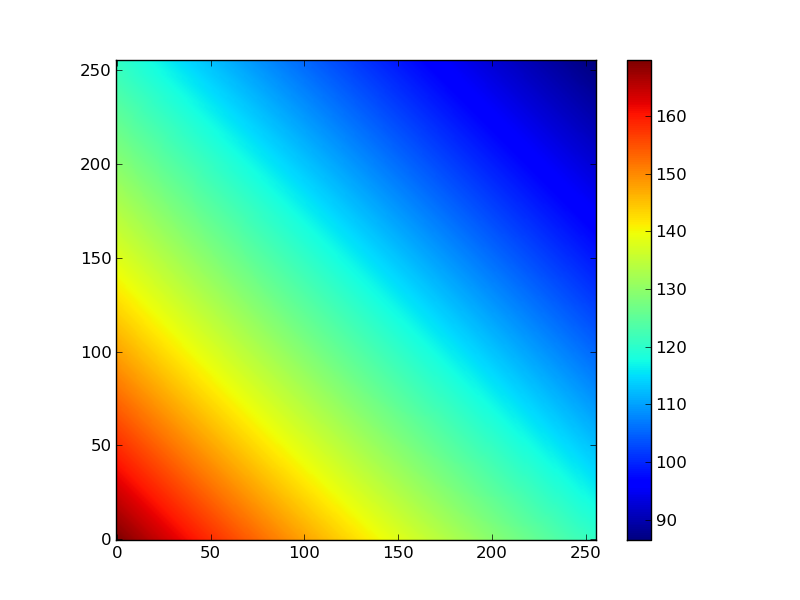}
  \includegraphics[trim=30 10 30 10,clip,width=0.4\textwidth]{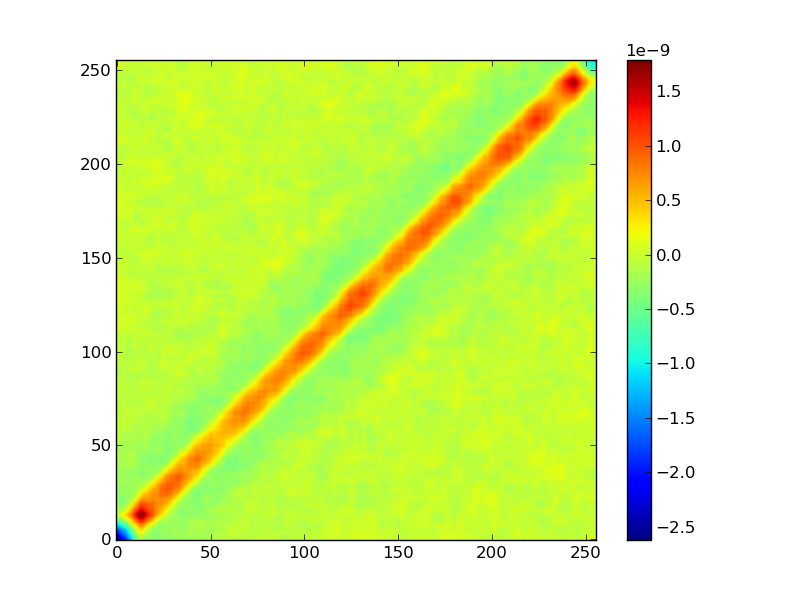} \\
  \includegraphics[trim=30 10 30 10,clip,width=0.4\textwidth]{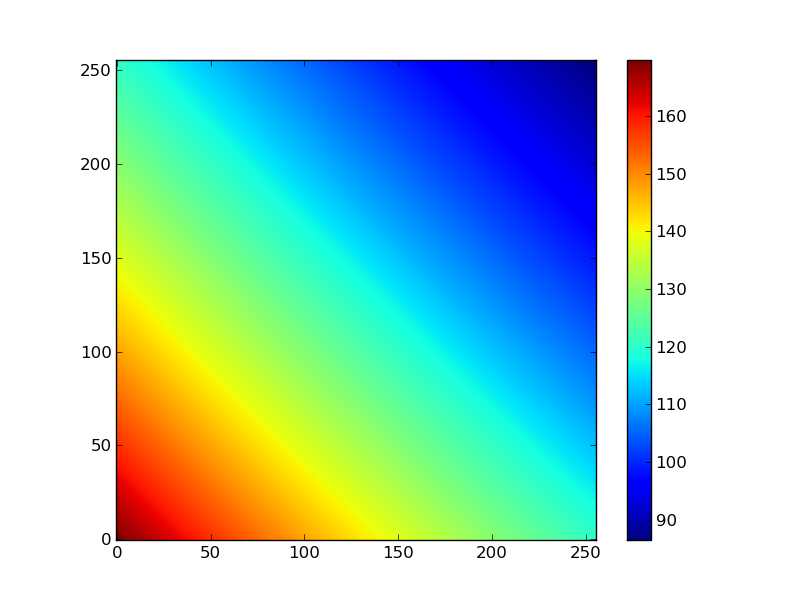}
  \includegraphics[trim=30 10 30 10,clip,width=0.4\textwidth]{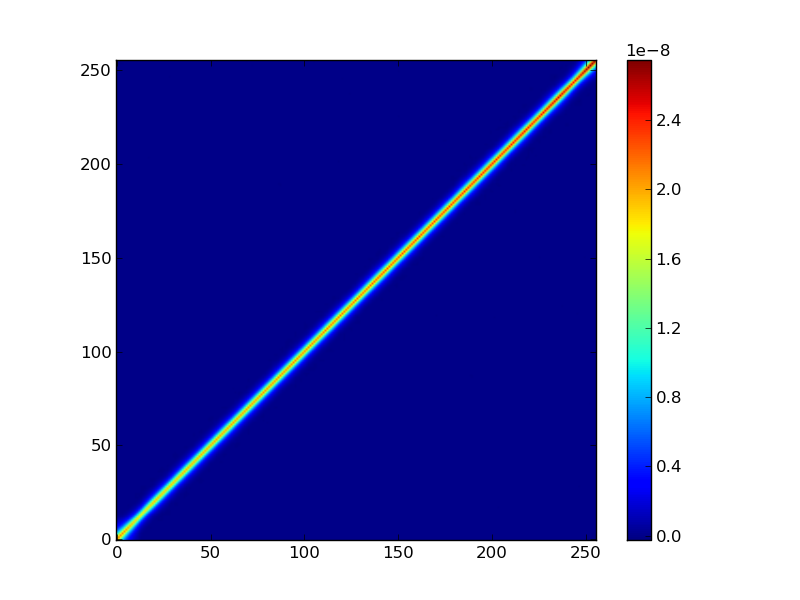}
  \caption{Frequency correlation matrix of data $\mat{R}$ (upper left), the
    recovered low-rank matrix $L$ (lower left) and the sparse matrix
    $S$ (lower right), and the difference of $\mat{R}_{\text{HI}} -
    S$ (upper right).}
  \label{fig:dec}
\end{figure*}

We use simulated sky map data to demonstrate the application of
the RPCA method for the separation of the foregrounds and the
21~cm signal. The real foreground may consist of many different physical components, 
such as synchrotron radiation, free-free emission, and emission by dusts. 
Here for demonstration we only include the two main components at low frequencies, 
i.e. the galactic synchrotron radiation and the extragalactic radio point sources. 
For this simulation, we use the Cosmology in the Radio Band 
(CORA)~\footnote{\url{https://github.com/radiocosmology/cora}}
\cite{2014ApJ...781...57S,2015PhRvD..91h3514S} package,  which  
simulates sky emission including galactic and extragalactic foregrounds,
with some extra utility codes for dealing with Healpix maps and spherical co-ordinates.
We consider a case which is inspired by the Tianlai experiment \cite{2012IJMPS..12..256C}
whose pathfinder works at the frequency range $700-800 \MHz$.
 We generate the sky maps in the frequency range $700-800$~MHz, with 256 equally
spaced frequency bins ($\Delta \nu \sim 0.39$~MHz), and the whole
sky is divided using the HEALPix pixelization scheme
\cite{2005ApJ...622..759G} with $n_{\text{side}} = 256$, which
corresponds to an angular resolution of $\sim 13.7$~arcmin.
To take the effect of frequency dependent beam into account, 
we first convolve the generated sky maps with a
symmetric circular frequency dependent Gaussian beam with 
$\theta=1.22 \lambda / D$, where $\lambda$ is the observing
wavelength, and $D$ is the diameter of the telescope. We take
$D=100$m, corresponding to the optimal size for the mid-redshift 21cm
intensity mapping experiment  \cite{2008PhRvL.100i1303C,2010ApJ...721..164S,2012A&A...540A.129A}, 
and also the size of current largest fully steerable telescope, 
such as the Green Bank Telescope (GBT) which is conducting intensity mapping 
observations \cite{2010Natur.466..463C,2013ApJ...763L..20M,2013MNRAS.434L..46S}.
For this size the beam width is $\sim 16.8$~arcmin at 750 MHz, which roughly matches 
the resolution of the map. The details of the simulation
is given in Appendix~\ref{S:gen}. We tested our
procedure by making 10 different realizations of the random field,
and found similar results in each case. One of these
is shown in \autoref{fig:maps}, at the central frequency of 750 MHz.

We use Algorithm \autoref{alg:RPCA_ALM} to solve the PCP problem, with a
fixed regularization parameter $\lambda = 1 / \sqrt{N_{\nu}}$, where
$N_{\nu}$ is the number of rows (or columns) of the $\nu - \nu'$
covariance matrix $\mat{R}$. After a few trials, we found that 
$\mu = 10^8$ works,  though it is not necessarily the optimal value. We
terminate the algorithm when $\norm{\mat{R} - L - S}_{F} \leq \delta \,
\norm{\mat{R}}_{F}$ with $\delta = 10^{-14}$. The algorithm
converges quickly with this choice of  $\lambda$ and $\mu$, it
takes less than 100 iterations to get the required precision.
The frequency covariance matrix $\mat{R}$ is then successfully decomposed 
into a low-rank component $L$, which is mainly the contribution of the 
foreground, and an almost diagonal sparse component $S$, which
represents the frequency covariance matrix of the 21~cm signal.  The
result is shown in \autoref{fig:dec}, from which we see  the
maximum element difference between $S$ and the input
$\mat{R}_{\text{HI}}$ is at least an order of magnitude lower than
the corresponding element of $\mat{R}_{\text{HI}}$, shown a very
accurate recovery of the 21~cm signal frequency covariance matrix
from the total signal.

\section{21cm Signal Recovery} \label{S:21cm}
Once we obtained the 21~cm $\nu - \nu'$ covariance matrix
$\mat{\hat{R}}_{\text{HI}}$ (the sparse matrix $S$)  and the foreground covariance matrix
$\mat{\hat{R}}_{f}$ (the low-rank matrix $L$), the 21cm signal  can be recovered from the data. 
A number of methods are available for this task,  some by using only $S$, and others may 
use both $L$ and $S$.  
For example, the Karhunen-Lo\`eve (K-L) transform method is one of the latter,
it seeks to find a linear transformation of the observed
data $\vec{x}' = \mat{P} \vec{x}$ such that by projecting onto the
transformation matrix $\mat{P}$ the signal covariance matrix $S$ becomes
diagonal and the noise (here the foreground) covariance matrix $L$
becomes the identity matrix $\mat{I}$:
\begin{align}
  S \to S' &= \mat{P} S \mat{P}^{T} = \mat{\Lambda}, \notag \\
  L \to L' &= \mat{P} L \mat{P}^{T} = \mat{I}, \label{eq:SL}
\end{align}
where $\mat{\Lambda}$ is a diagonal matrix, and $\mat{I}$ is the identity
matrix.  The sub-space with low foreground contamination can then be 
identified as those with larger values of diagonal elements of $\mat{\Lambda}$.
Mathematically, this transformation can be found by solving the generalized
eigenvalue problem $S \vec{x} = \lambda L \vec{x}$, or written in matrix
form $S \mat{X} = L \mat{X} \mat{\Lambda}$. This gives a set of eigenvectors
$\vec{x}$ (or $\mat{X}$), and corresponding eigenvalues $\lambda$ (or
$\mat{\Lambda}$), then $\mat{X} = \mat{P}^{T}$, i.e. the
transformation matrix $\mat{P}$ consists each eigenvectors $\vec{x}$ as
a row. To recover the 21~cm signal,  select modes with eigenvalue (which is the signal-to-foreground power)
 greater than a certain threshold. Define the matrix $\mat{P}_{s}$ which contains
only the rows from $\mat{P}$ corresponding to eigenvalues greater than
the threshold $s$, the 21~cm signal is then recovered approximately by 
 $\vec{\hat{s}}= \mat{P}_{s} \vec{x}$.

This 21~cm signal recovery method depends on the joint transformation of
$L$ and $S$, the inaccuracy in either $L$ and $S$ may lead to large errors.
Another potential problem of this method is that for the generalized eigenvalue problem to
have a solution, $L$ must be symmetric positive definite, but there
is no guarantee for this in the RPCA decomposition, and
the solution would fail if $L$ is not positive definite. 
For these reasons, here we choose to uses only the $S$ matrix, which is generally
more robust than methods which depend on both $L$ and $S$. 
Note that to do RPCA decomposition, both the 
sparsity of $S$ and low-rank of $L$ and needed, but to recover the 21~cm signal, 
$S$ is sufficient.

In this paper we use the generalized Internal Linear Combination (ILC) method
\cite{2011MNRAS.418..467R} to recover the 21~cm signal. The ILC
component separation method has been extensively used to extract the
cosmic microwave background (CMB) from the WMAP multi-frequency
data
\cite{2003ApJS..148...97B,2003PhRvD..68l3523T,2009A&A...493..835D}. However,
here we will follow the method and symbol notations presented in
\cite{2016MNRAS.456.2749O} because which also deals with HI signal
extraction, thus closer to our work.

We have already mentioned that foreground components are correlated
over frequencies, so we expect that the foregrounds signal can
be represented as a linear combination of a finite number $m$
of independent templates, which do not necessarily represent  physical components. In other
words, we try to capture all foreground contributions as resulting
from $m$ (unphysical) templates, so $m$ is just the effective
dimension of the foregrounds subspace. While the
21~cm signal $\vec{s}$ is only correlated over adjacent frequencies, and 
can be represented as the linear combination of $N_{\nu} - m$ independent (unphysical)
templates $\vec{t}$,
\begin{equation} \label{eq:st}
  \vec{s} = \mat{S} \vec{t},
\end{equation}
where $\mat{S}$ is a $N_{\nu} \times (N_{\nu} - m)$ mixing matrix
giving the contribution from the templates to the HI emission in
each frequency channel. 
Note that we are not interested in recovering $\vec{t}$, our goal is
just to use it to explore the subspace of the HI
signal and recover $\vec{s}$. Using this expression, we can write
the 21~cm $\nu - \nu'$ covariance matrix as
\begin{equation} \label{eq:RHI}
  \mat{R}_{\text{HI}} = \mat{S} \mat{R}_{t} \mat{S}^{T},
\end{equation}
where $\mat{R}_{t} = \langle \vec{t} \vec{t}^{T} \rangle$ is the
$(N_{\nu} - m) \times (N_{\nu} - m)$ covariance matrix of the
templates $\vec{t}$.
The generalized ILC method estimates
the signal $\vec{s}$ as a linear combination of the total signal
$\vec{x}$ as
\begin{equation} \label{eq:sest}
  \hat{\vec{s}} = \mat{W} \vec{x},
\end{equation}
where $\mat{W}$ is the $N_{\nu} \times N_{\nu}$ ILC weight matrix,
which have unit response to the 21~cm signal, while minimizing the
total variance of the vector estimate $\hat{\vec{s}}$. The optimal
weighting matrix of the generalized multi-dimensional ILC 
can be written as \cite{2011MNRAS.418..467R}
\begin{equation} \label{eq:W}
  \mat{W} = \mat{S} (\mat{S}^{T} \mat{R}^{-1} \mat{S})^{-1}
  \mat{S}^{T} \mat{R}^{-1}.
\end{equation}
We can see from Eq.~(\ref{eq:W}) that $\mat{W}$ is
invariant under the transformation $\mat{S} \to \mat{S}\mat{T}$ for
any invertible matrix $\mat{T}$, so
$\vec{t}$ can also be replaced by any other linear combination
$\mat{T} \vec{t}$. We apply the following
transformation to the total signal $\vec{x}$ using the estimated 21~cm $\nu - \nu'$ 
covariance matrix $\mat{\hat{R}}_{\text{HI}}$,
\begin{equation} \label{eq:Rx}
  \vec{x} \to  \mat{\hat{R}}_{\text{HI}}^{-1/2} \vec{x},
\end{equation}
where  the matrix $\mat{\hat{R}}_{\text{HI}}^{-1/2}$ is defined to
satisfy $\mat{\hat{R}}_{\text{HI}}^{-1/2} \mat{\hat{R}}_{\text{HI}}
\mat{\hat{R}}_{\text{HI}}^{-1/2} = \mat{I}$, and can be calculated
by eigen-decomposition: 
$\mat{\hat{R}}_{\text{HI}}= \mat{U} \mat{\Lambda} \mat{U}^{T}$, then
$\mat{\hat{R}}_{\text{HI}}^{-1/2} = \mat{U} \mat{\Lambda}^{-1/2} \mat{U}^{T}$.

\begin{figure}[htbp]
  \centering
  \includegraphics[trim=20 20 30 20,clip,width=0.45\textwidth]{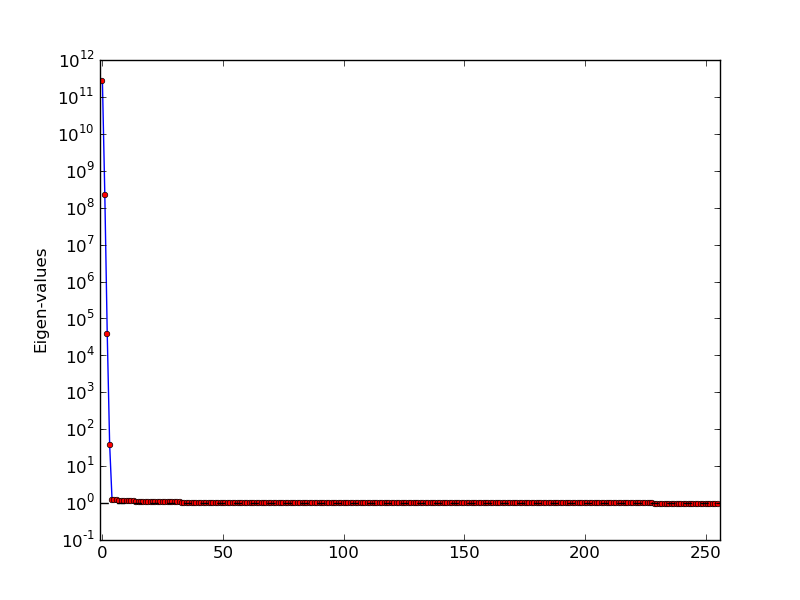}
  \includegraphics[trim=20 20 30 20,clip,width=0.45\textwidth]{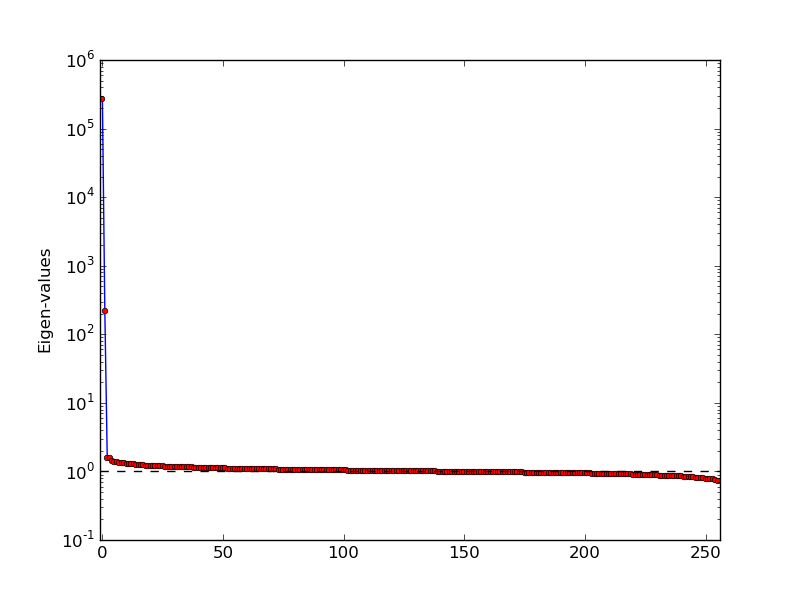}
  \caption{Eigen-values of the covariance matrix
    $\mat{C}=\mat{\hat{R}}_{\text{HI}}^{-1/2} \mat{R}
    \mat{\hat{R}}_{\text{HI}}^{-1/2}$. Top is for the parameters
    setting of this simulation, four eigen-values are significantly
    larger; bottom is for the case when the 21~cm signal is 1000
    times larger, now there is two significantly lager eigen-values.}
  \label{fig:eigval}
\end{figure}

\begin{table*}[htbp]
  \centering
    \caption{The largest 20 eigenvalues. }
  \begin{ruledtabular}
  \begin{tabular}{ccccc}
    $2.680 \times 10^{11}$ & $2.187 \times 10^8$ & $3.751 \times 10^4$ & $3.824 \times 10^1$ & 1.201 \\
    \hline
    1.188 & 1.181 & 1.169 & 1.164 & 1.147 \\
    \hline
    1.135 & 1.129 & 1.119 & 1.108 & 1.099 \\
    \hline
    1.089 & 1.082 & 1.073 & 1.063& 1.055 \\
  \end{tabular}
\end{ruledtabular}
  \label{tab:eig20}
\end{table*}

The transformed quantity $\mat{\hat{R}}_{\text{HI}}^{-1/2} \vec{x}$
will have a covariance
\begin{eqnarray} \label{eq:RRR}
\mat{C} &\equiv&  \mat{\hat{R}}_{\text{HI}}^{-1/2} \mat{R}   \mat{\hat{R}}_{\text{HI}}^{-1/2} \nonumber\\
&=&  \mat{\hat{R}}_{\text{HI}}^{-1/2} \mat{R}_{f} \mat{\hat{R}}_{\text{HI}}^{-1/2} +
  \mat{\hat{R}}_{\text{HI}}^{-1/2} \mat{R}_{\text{HI}}  \mat{\hat{R}}_{\text{HI}}^{-1/2}.
\end{eqnarray}
The estimated HI $\nu - \nu'$ covariance matrix $\mat{\hat{R}}_{\text{HI}}$ is close to the real
HI covariance matrix $\mat{R}_{\text{HI}}$, so the last term in
Eq.~(\ref{eq:RRR}) will be close to the identity matrix $\mat{I}$.
Now if we make an eigen-decomposition of the left covariance matrix
of Eq.~(\ref{eq:RRR}), we have
\begin{equation} \label{eq:eig}
 \mat{C}= [\mat{U}_{F} \mat{U}_{S}]
  \times \begin{bmatrix} \lambda_1 + 1 & & & \\ & \cdots & &  \\ & & \lambda_m +1 & \\ & & & 
  \mat{\tilde{I}}  \\
\end{bmatrix} \, \times  \, \begin{bmatrix} \mat{U}_F^T  \\ \mat{U}_S^T \end{bmatrix},
\end{equation}
where $\mat{\tilde{I}}$ is used to denote the appropriate identity
sub-matrix. From this we see that the eigenvalues
of the covariance matrix $\mat{C}=\mat{\hat{R}}_{\text{HI}}^{-1/2} \mat{R}
\mat{\hat{R}}_{\text{HI}}^{-1/2}$ that are nearly unity
contains the power of the HI 21~cm signal, with the
corresponding eigenvectors spanning the HI subspace. This is
equivalent to say that the eigenvalues significantly greater than unity
contains the foreground components, and the number of
these eigenvalues determines $m$, the dimension of foreground subspace. 

We can extract the subspace corresponding to the HI 21~cm signal as
\begin{equation} \label{eq:HIsub}
  \mat{\hat{R}}_{\text{HI}}^{-1/2} \mat{R}_{\text{HI}}
  \mat{\hat{R}}_{\text{HI}}^{-1/2} = \mat{U}_{S} \, \mat{\tilde{I}} \, \mat{U}_{S}^{T},
\end{equation}
which can be re-written as
\begin{equation} \label{eq:RHIest}
  \mat{R}_{\text{HI}} = \mat{\hat{R}}_{\text{HI}}^{1/2}
  \mat{U}_{S} \, \mat{\tilde{I}} \, \mat{U}_{S}^{T} \mat{\hat{R}}_{\text{HI}}^{1/2}.
\end{equation}
This is just the $N_{\nu} \times N_{\nu}$ HI $\nu - \nu'$ covariance
matrix projected onto the $(N_{\nu} - m)$-dimensional HI subspace
spanned by the subset of eigenvectors collected in matrix
$\mat{U}_{S}$. Compare Eq.~(\ref{eq:RHIest}) with Eq.~(\ref{eq:RHI}),
and using the fact that the template $\vec{t}$ can be replaced by
any linear combination $\mat{T} \vec{t}$ provided $\mat{T}$ is
invertible, we can choose $\mat{T}$ to be the orthogonal transform
matrix of $\mat{R}_{t}$ to re-express Eq.~(\ref{eq:RHI}) as
\begin{equation} \label{eq:RHI1}
  \mat{R}_{\text{HI}} = \mat{S} \mat{T} \mat{D}_{t} \mat{T}^{T} \mat{S}^{T},
\end{equation}
where $\mat{D}_{t}$ is the diagonal matrix of the orthogonal
transform $\mat{R}_{t} = \mat{T} \mat{D}_{t} \mat{T}^{T}$. Now we
can  see that
\begin{equation} \label{eq:ST}
  \mat{S} \mat{T} = \mat{\hat{R}}_{\text{HI}}^{1/2} \mat{U}_{S}
\end{equation}
up to an scaling factor which could be absorbed into
$\mat{T}$. Since $\mat{W}$ is invariant under the transform of
$\mat{S} \to \mat{S}\mat{T}$ for any invertible matrix $\mat{T}$, we
get the estimate of the mixing matrix of the 21~cm signal as
\begin{equation} \label{eq:Sest}
  \mat{\hat{S}} = \mat{\hat{R}}_{\text{HI}}^{1/2} \mat{U}_{S}.
\end{equation}

Now we can substitute Eq.~(\ref{eq:Sest}) into Eq.~(\ref{eq:W}), then from
Eq.~(\ref{eq:sest}) we can get the recovered HI 21~cm signal.
We show that the eigenvalues of the covariance matrix $\mat{C}$
 for the simulated data in
\autoref{fig:eigval} and the largest 20 eigen-values in \autoref{tab:eig20}.

\begin{figure}[htbp]
 \centering
 \includegraphics[trim=10 10 10 10,clip,width=0.45\textwidth]{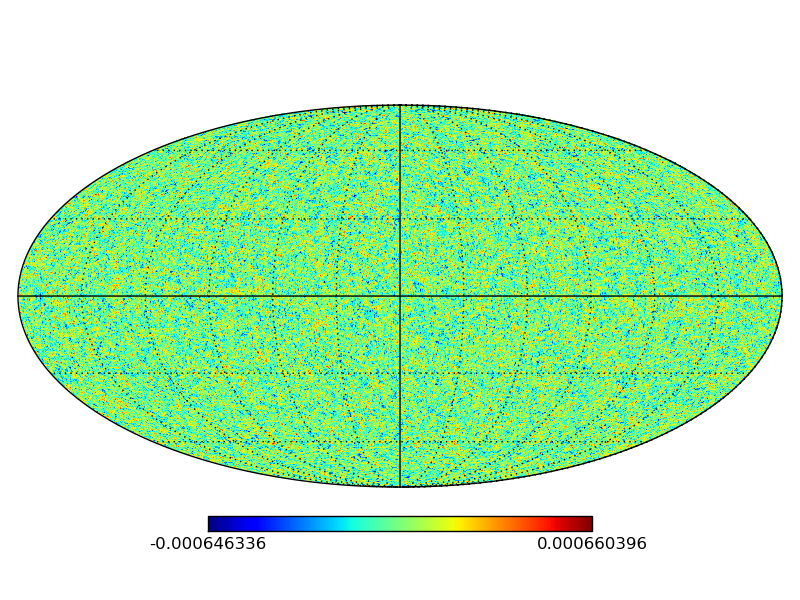}
 \includegraphics[trim=10 10 10 10,clip,width=0.45\textwidth]{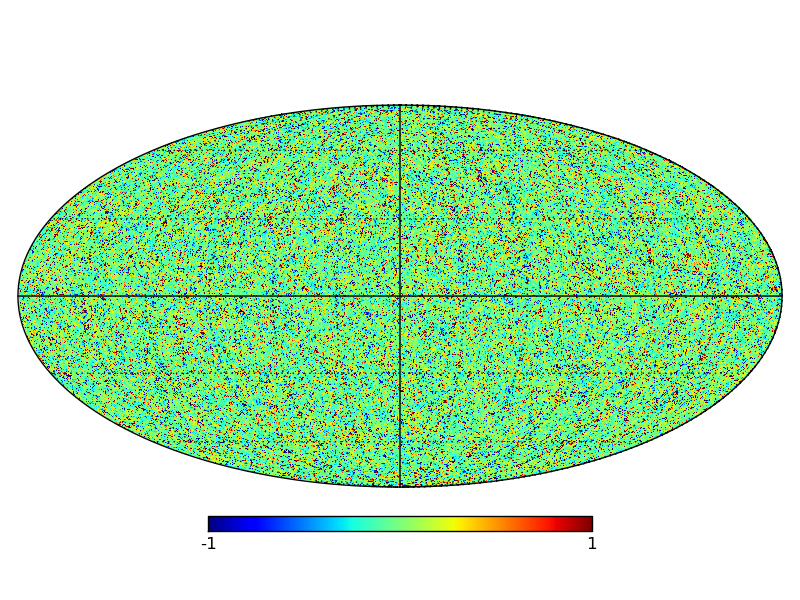}
 \caption{HI 21~cm signal recovery result for a threshold
   2.0. Subfigures are: (a) the recovered 21~cm signal, (b) the
   relative difference of the input 21~cm signal and the recovered
   21~cm signal.}
 \label{fig:rec2.0}
\end{figure}

\begin{figure}[htbp]
 \centering
 \includegraphics[trim=5 10 30 10,clip,width=0.45\textwidth]{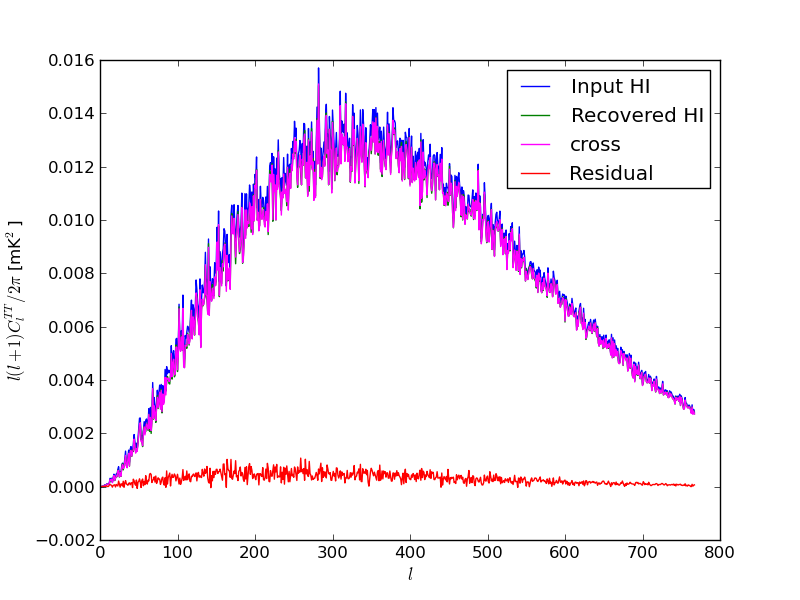}
 \includegraphics[trim=5 10 30 10,clip,width=0.45\textwidth]{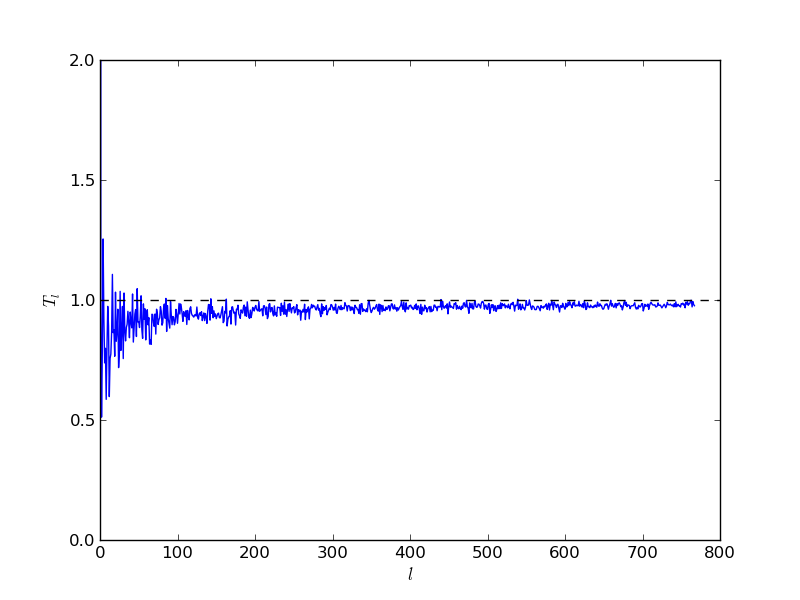}
 \caption{The normalized angular power spectrum $l(l + 1) C_{l} /2\pi$ 
 of the recovered HI 21~cm signal for a threshold 2.0 (top), where
    the curve labeled ``cross'' is the cross angular power spectrum
    of the input 21~cm signal and the recovered 21~cm signal, and
    the transfer function $T_{l} = C_{l}^{\text{input}} / C_{l}^{\text{recovered}}$ (bottom).}
 \label{fig:cl2.0}
\end{figure}

From Eq.~(\ref{eq:eig}) we see the eigenvalues of $\mat{C}$ are $1+\lambda_m$, so 
if we select $\lambda_m >1$ or the eigenvalues of $\mat{C}$ greater than 2,  
the subspace of the eigenvectors would correspond to the low rank matrix $\mat{L}$,
so in fact there is no arbitrary choice here. 
From \autoref{fig:eigval} and \autoref{tab:eig20}, we see that in
this simulation there are four large eigenvalues, all other eigenvalues are
significantly smaller. As we will see, this number is also the optimal number of eigen-modes
that should be subtracted in the classic PCA method, which actually
reflects the fact that our foreground model has four principal
components that dominates the 21~cm signal. We have also made tests to 
check this correspondence. For example, if we artificially increase
the 21~cm signal by 1,000 times while keeping the foregrounds
unchanged, now the signal actually dominates the smaller two
principal components of the foregrounds, and we see there is
only two eigenvalues that are significantly larger as shown in the
bottom of \autoref{fig:eigval}. The GILC method
can quite automatically and robustly detected the dimensionality of
the foreground subspace, which is an additional advantage of our
method against the classic PCA method we will discuss next.

We show in Fig.\ref{fig:rec2.0} (a) the recovered 21~cm signal, (b)
the relative difference of the input 21~cm signal and the recovered
21~cm $(\vec{x}^{\text{recovered}} - \vec{x}^{\text{input}}) /
(\bar{T}_{b} + \vec{x}^{\text{input}})$, where $\bar{T}_{b}$ is the mean
temperature of the 21~cm signal computed according to \autoref{eq:Tbz}. For fully 
accurate recover, the relative difference is 0. 
We can see from the figure that the signal is generally well recovered, as most points
of the map the relative error is nearly zero. The residue deviations 
are randomly distributed over the sky. 

We can also check the recovery in spherical harmonic space, which is also what we are ultimately
interested in cosmology. The
cross angular power spectrum between the input and recovered map is defined as 
$$C_{l}^{\text{cross}} = \frac{1}{2l + 1} \sum_{m}\frac{1}{2}(a_{lm}^{\text{input}} a_{lm}^{\text{recover}, *} +a_{lm}^{\text{input}, *} a_{lm}^{\text{recover}}).$$
where $a_{lm}^{\text{input}}, a_{lm}^{\text{recover}}$ are the spherical harmonic expansion 
coefficients for the input and recovered map respectively. The auto power spectrum for the 
input map, recovered map and the cross power  
are shown in \autoref{fig:cl2.0}. These power spectra
almost coincides with each other, and the residue difference is very small. 

We can also compute the angular power transfer function $T_{l} = C_{l}^{\text{input}} / C_{l}
^{\text{recovered}}$, which is shown as the bottom panel of Fig.\ref{fig:cl2.0}, 
which is nearly unity for the interested $\ell$ range, though the error is larger at small $l$ due 
to cosmic variance.

\section{Comparison with the Classic PCA} \label{S:comp}

In contrast to the classic PCA/SVD foreground subtraction method
which only uses the low-rank structure and characteristics of the
foregrounds implicitly, the RPCA method presented in
this paper exploits the additional characteristics of the HI 21~cm
signal frequency covariance matrix as well, which has a very
sparse structure, as discussed in \autoref{S:fc}, only the elements
along and near the main diagonal are non-zero. This improves the 
effectiveness of foreground subtraction, and is free from the 21 cm signal loss
problem encountered in the classic PCA/SVD method. 

To compare with the classic PCA method, we follow the method presented in
\cite{2015MNRAS.447..400A,2015MNRAS.454.3240B}. 
Diagonalize the frequency covariance matrix $\mat{R}$ of the full data set
with an eigen-decomposition, 
\begin{equation} \label{eq:ed}
  \mat{U}^{T} \mat{R} \mat{U} = \mat{\Lambda} \equiv \text{diag} \{
  \lambda_{1}, \cdots, \lambda_{N_{\nu}} \},
\end{equation}
The magnitude of $\lambda_{i}$ gives the variance of the corresponding eigen-mode, 
each eigenvalue measures the contribution of its
corresponding eigenvector to the total sky variance.  As the
foreground components dominate the full data overwhelmingly, we
expect they would occupy the modes with the largest eigenvalues, so we
pick a number of the largest eigenvalues $\lambda_{1} \ge \lambda_{2}
\ge \cdots \lambda_{k} \ge 0$ to construct a matrix $\Lambda_{f} =
\text{diag} \{ \lambda_{1}, \cdots, \lambda_{k}, 0, \cdots, 0 \}$, and use their corresponding
eigenvectors to build a matrix $\mat{U}_{f}$, then use this matrix
to extract the foregrounds from the full data as 
$\hat{\vec{f}} = \mat{U}_{f} \mat{U}_{f}^{T} \vec{x}.$
Its covariance matrix is
$ \hat{\mat{R}}_{f} = \frac{1}{N_p} \langle \hat{\vec{f}} \hat{\vec{f}}^{T} \rangle 
= \mat{U}_{f} \mat{\Lambda}_{f} \mat{U}_{f}^{T},$
This is the solution of the classical PCA problem Eq.~(\ref{eq:cp}) with
$M$ being the full data covariance matrix $\mat{R}$ and $k$ being
the number of largest eigenvalues corresponding to the contribution
of foregrounds. The  21~cm signal can then be reconstructed by subtracting the
foregrounds from the total data with
 $ \hat{\vec{s}} = \vec{x} - \hat{\vec{f}}$.

\begin{figure}[htbp]
  \centering
  \includegraphics[trim=20 20 30 20,clip,width=0.45\textwidth]{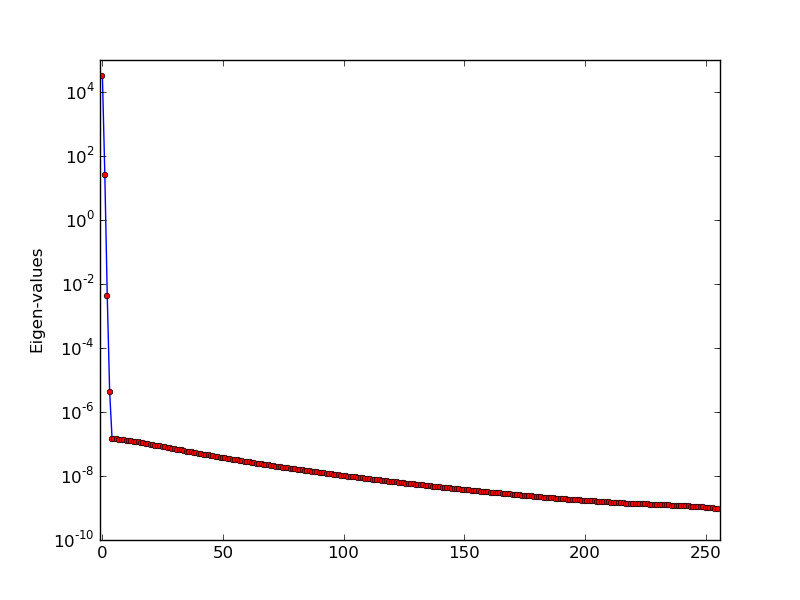}
  \caption{Eigen-values of the covariance matrix $\mat{R}$.}
  \label{fig:pca-eig}
\end{figure}

For our simulated data, we show the eigenvalues in \autoref{fig:pca-eig}.
We see  that there are four eigenvalues
which are significantly larger than others, the other eigenvalues
decay slowly, which is different from that in \autoref{fig:eigval},
where aside from the largest four eigenvalues, all others are quite
close to 1, because by doing the transformation \autoref{eq:Rx}, we
have made the covariance matrix corresponding to the signal subspace
close to identity. By subtracting the first 4
eigenmodes we obtain the reconstructed HI 21~cm signal.

\begin{figure}[htbp]
  \centering
  \includegraphics[trim=0 10 10 10,clip,width=0.45\textwidth]{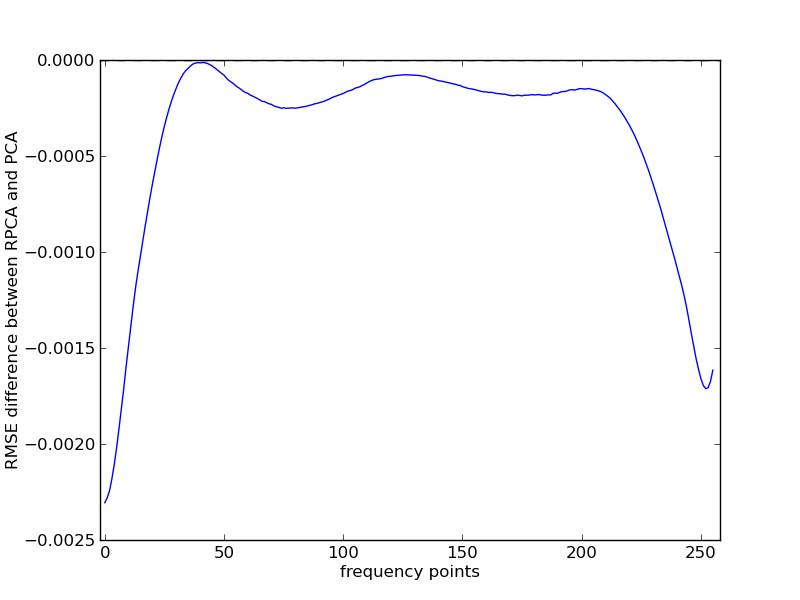}
  \includegraphics[trim=0 10 10 10,clip,width=0.45\textwidth]{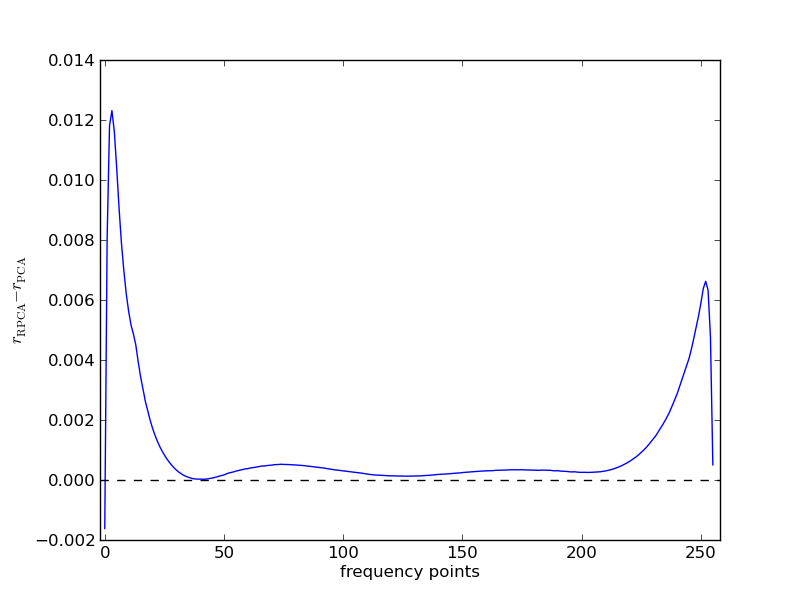}
  \caption{Difference of RMSE (top) and Pearson correlation coefficient $r$
    (bottom) between the RPCA+GILC method and the classical
    PCA method for all frequency points.}
  \label{fig:diff}
\end{figure}

To compare the robust and classical PCA results, we introduce  
two quantitative aspects: the Root-Mean-Square Error (RMSE) and the Pearson correlation
coefficient $r$. The former is defined as
$$ \text{RMSE} = \sqrt{\sum\nolimits_{i}(x_{i}^{\text{input}} -
  x_{i}^{\text{recovered}})^{2}}, $$
which reflects the total disagreement between the recovered 21~cm
signal $\vec{x}^{\text{recovered}}$ and the input 21~cm signal
$\vec{x}^{\text{input}}$; the latter is defined as
$$ r = \frac{\sum_{i}(x_{i}^{\text{input}} -
  \bar{x}^{\text{input}})(x_{i}^{\text{recovered}} -
  \bar{x}^{\text{recovered}})}{\sqrt{\sum_{i}(x_{i}^{\text{input}}
    - \bar{x}^{\text{input}})^{2}} \sqrt{\sum_{i}(x_{i}^{\text{recovered}}
    - \bar{x}^{\text{recovered}})^{2}}}, $$
which is a measure of the linear correlation between the recovered 21~cm
signal $\vec{x}^{\text{recovered}}$ and the input 21~cm signal
$\vec{x}^{\text{input}}$, where $\bar{x}^{\text{recovered}}$ and
$\bar{x}^{\text{input}}$ are the mean of
$\vec{x}^{\text{recovered}}$ and $\vec{x}^{\text{input}}$,
respectively. 

We show the difference of RMSE and $r$ between
the two methods in \autoref{fig:diff} for each frequency point, where we see the 21~cm signal
recovered by the RPCA + GILC method is consistently better
than that recovered by the classical PCA method in the whole
frequency band, though in this case the difference is not large. 
This is probably because the foreground in our simulation can be well
approximated by a rank-4 matrix, so the classical PCA extraction is already pretty good.

%

\begin{figure}[htbp]
 \centering
 \includegraphics[trim=15 10 30 10,clip,width=0.45\textwidth]{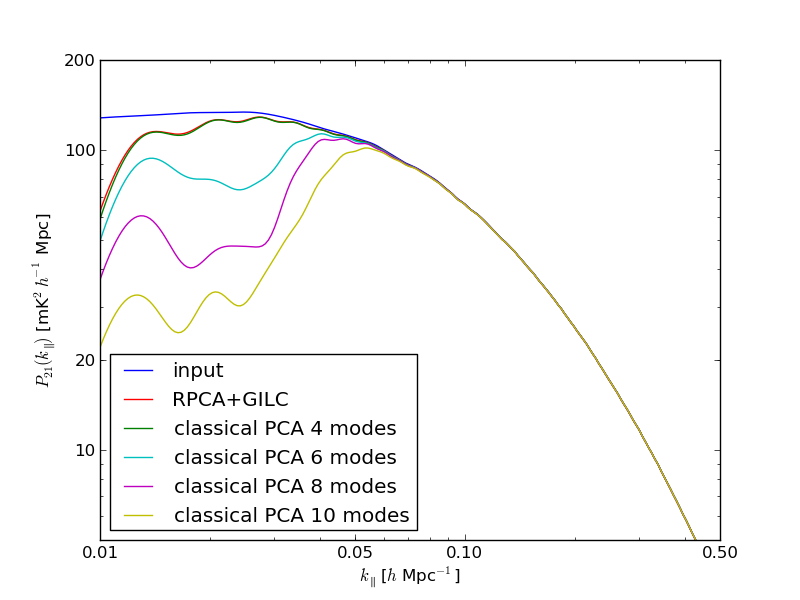}
 \caption{The line of sight temperature power spectrum
   $P_{21}(k_{\parallel})$ of the recovered 21~cm signal.}
 \label{fig:pk}
\end{figure}

We plot the line of sight temperature power
spectrum $P_{21}(k_{\parallel})$ of the recovered signal  
by the RPCA  method and by the classical PCA method 
with different number of PCA modes being subtracted in
\autoref{fig:pk}. Both performs well on small scales (large $k$). On the large scales (small $k$), 
the performance of both methods get worse as here the signal is more similar to the smooth foregrounds. 
The oscillating structure in the recovered $P_{21}(k_{\parallel})$ is due to the
limited bandwidth (100~MHz) we have used. However, in the classical PCA the number of modes to 
be subtracted is more uncertain, as the decay of the eigenvalues is slower near the floor, as
shown in Fig.~\ref{fig:pca-eig}, making it harder to decide how many modes should be subtracted. 
If the number of PCA modes being subtracted is incorrectly set (in this figure more than optimal 4 modes), 
the recovery performance of the classical PCA methods would become significantly worse as 
more modes with higher signal-to-foreground ratio are being subtracted, which causes
the signal loss problem.  For the recovery of the power spectrum, this signal-loss could be partially
compensated by computing the signal loss transfer function from numerical 
simulations \cite{2013ApJ...763L..20M,2013MNRAS.434L..46S}, but it is still a tricky and time-consuming process.
In the RPCA method, at least in principle, the separation of the 21cm signal and the foreground is automatic, and as 
shown in Fig.~\ref{fig:eigval} the eigenvalues are more distinctly separated.

However, despite the advantage of the RPCA method over the classic PCA method, 
in the recovered 21~cm signal and power spectrum the difference is not large. 
While we showed only one example above, we have 
tried various different realizations and different  mean signal-to-foreground ratios, the results are generally 
similar to what we get here, the RPCA result is only slightly better than the best result of the
classical PCA with optimal subtraction of modes.  A possible reason for this 
is that the simulated foreground and beam models are so simple that 
the classical PCA already subtracted most of the foregrounds in the simulation. 
In practice, however, a much larger number of principal components
must be subtracted from the the real observation data  using the classical PCA. For example,
in Ref. \cite{2013ApJ...763L..20M}, after many trials the first 20 modes are subtracted
  to get the 21~cm map. The signal loss would be quite severe, one needs to 
  use simulation of mock samples to measure the power spectrum transfer function in the
  SVD  in order to compensate for the signal loss. In that case 
  the advantage of the RPCA method would be more significant. Analysis of the real
  data, however, require to deal with a number of practical issues. In the present paper we 
  will concentrate on the algorithm itself, while the application to the observational data would
  be investigated in a subsequent study.

\begin{figure}[htbp]
  \centering
  \includegraphics[trim=25 10 50 20,clip,width=0.45\textwidth]{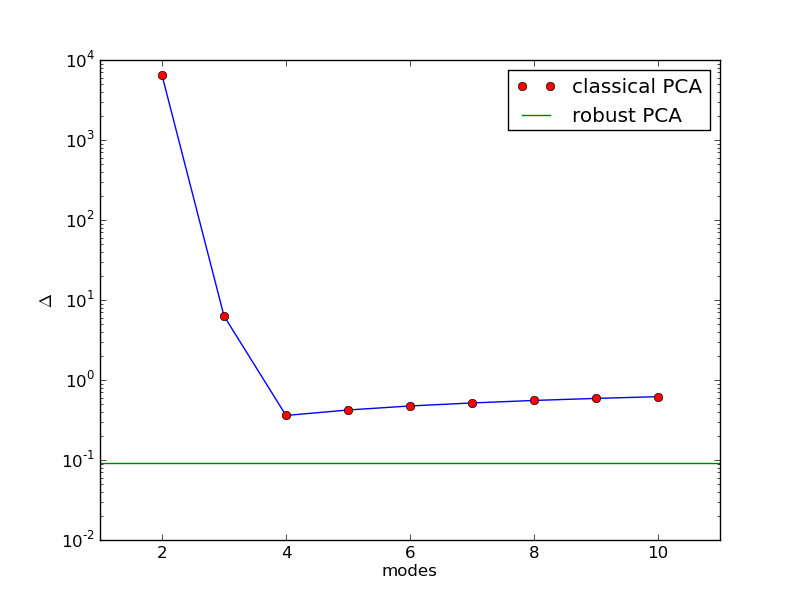}
  \caption{The relative recovery error of the 21~cm signal for the two method: red dots for the classical PCA method, green line for the RPCA method. $x$-axis is the number of modes subtracted by the classical PCA method.}
  \label{fig:loss}
\end{figure}

By utilizing both the low-rank property of the foreground frequency covariance
and the sparsity of the 21~cm signal frequency covariance, the
RPCA method can be used to subtract the foreground without losing the 
signal, which  was encountered in the classic PCA method. 
However, the GILC reconstruction process may not be the best reconstruction method and 
also causes some signal loss too. To make a fair comparison of the RPCA 
with the classical PCA without being affected by the additional GILC part, we define a quantity 
$\Delta = \norm{\hat{\mat{R}}_{\text{HI}} - \mat{R}_{\text{HI}}}_{F} / \norm{\mat{R}_{\text{HI}}}_{F}$ 
 as the relative error between
the recovered 21~cm signal frequency covariance matrix $\hat{\mat{R}}_{\text{HI}}$
and the input 21~cm signal frequency covariance matrix $\mat{R}_{\text{HI}}$.
We show the relative error
$\Delta$ in \autoref{fig:loss} for the RPCA method and also the
classical PCA method with different number of PCA modes subtracted,
where $\hat{\mat{R}}_{\text{HI}}$ is the sparse component $S$ for the 
RPCA method, and $\langle \hat{\vec{s}} \hat{\vec{s}}^{T} \rangle /
N_{p}$ for the classical PCA method
where $\hat{\vec{s}}$ is the recovered 21~cm signal. 
In \autoref{fig:loss}, we see a clear turning point at $N=4$
for the number of subtracted modes for the classical PCA method. This shows
that the relative error decreases as we subtract the large PCA modes initially, 
but then the relative error begin to increase again as we subtract
more modes, indicating more and more severe 21~cm
signal loss during the classical PCA mode subtracting process. The relative
recovery error of the RPCA method is  about an
order of magnitude lower than even the lowest relative error of the
classical PCA method, indicating a better foreground subtraction
and 21~cm signal recovery result. 

We see that the RPCA method and the PCP algorithm are actually quite
automatic in the sense that less tunable parameters are needed for
the user to provide or tune, the regularization parameter $\lambda$
can be taken as $\sqrt{\max{(m, n)}}$ and works well
in almost all cases, the user only need to provide an appropriate
threshold $\delta$ for the iteration stopping criteria, Algorithm
\autoref{alg:RPCA_ALM} automatically gets the matrices $L$ and $S$
satisfying the required low-rank and sparsity condition, the
rank $k$ of $L$ and the number of non-zero elements and their
support do not need to be known as a prior or to be determined, while
for the classical PCA/SVD foreground subtraction, the user has to do
several trials to determine the
optimal number of principal components or SVD modes for a balance of
good 21~cm signal recovery result and minimum signal loss.

\section{Discussion} \label{S:dis}
We have shown by simulation that the RPCA method can efficiently
extract the HI 21~cm $\nu - \nu'$ covariance matrix  from the
observed data with better accuracy than the classical 
PCA method by utilizing the generic condition that 
the 21cm signal covariance matrix is sparse, while the foreground covariance matrix has low ranks. 
Algorithm \autoref{alg:RPCA_ALM} is very fast, 
only a relatively small number of iterations is needed to achieve good relative accuracy. 
For a $256 \times 256$ matrix decomposition used in this simulation, it
takes less than a minute on a personal computer to converge to a 
satisfactory error threshold. For higher pixel and frequency bin numbers, 
the speed of computation could be further accelerated by noting that instead of doing 
a full matrix SVD decomposition in each iteration, for calculating
the singular value thresholding operator $\mathcal{D}_\tau(X)$ (step 3 in
Algorithm \autoref{alg:RPCA_ALM}),  only those singular values that are
greater than the threshold $\tau$ and the corresponding singular
vectors are needed, so a partial SVD suffices \cite{Cai2010b}.  
Some algorithms and softwares can do this, 
e.g., PROPACK\footnote{\url{http://soi.stanford.edu/~rmunk/PROPACK/}} and the
modified version of LANSVD\footnote{\url{http://svt.stanford.edu/code.html}} which comes
with a threshold option to compute only those singular vectors with
singular values greater than a given threshold value $\tau > 0$.

In practice, artifacts are introduced in the sky maps reconstructed
from the observatories during the map-making process, due to
imperfections such as incomplete
uv-coverage, missing data, particularly bright source, numerical errors.
Usually the artifacts are not correlated with the HI 21~cm signal, 
though in some case may have correlations with the
foregrounds, and may be distinct from both the foregrounds 
and the HI 21~cm signal. After removing some most obvious ones, 
residue artifacts should be small relative to the diffuse foregrounds and extragalactic point
sources,  though still larger than that of the HI 21~cm signal. Due to the
complex structure and frequency dependence, its covariance matrix 
will not have a rank as low as that of the foregrounds, nor will it
be sparse like that of the HI 21~cm signal. In the presence of artifacts, an extension 
of the RPCA method presented above is to include an additional term,
 $$\mat{R} = L + S+ Z,$$ 
 where $Z$ is a dense, small perturbation, which can be solved via the convex
optimization problem called Stable Principle Component Pursuit (SPCP)
\cite{Zhou2010}:
\begin{equation} \label{eq:SPCP}
\min_{L,S} \; \norm{L}_{*} + \lambda \norm{S}_{1} \quad
\textup{s.t.} \quad \norm{\mat{R} - L - S}_{F} \leq \delta.
\end{equation}
Efficient algorithms for this problem are available
\cite{Aybat2014,Aybat2015}. 

The GILC method is used to recover the 21~cm signal from the observation data
using the 21~cm $\nu - \nu'$ covariance matrix extracted by the
RPCA method.  In this paper, the method is applied to the 
sky map as a whole, but in fact the foreground components  may vary over 
different sky areas or different angular scales.
Improvement can be made by employing the 
Generalized Needlet Internal Linear Combination (GNILC) method
\cite{2011MNRAS.418..467R,2016MNRAS.456.2749O}, which
worked in a needlet frame, the number of principal
components of the observed covariance matrix is estimated locally
both in space and in angular scale by using a wavelet (needlet)
decomposition of the observations. We will explore how this
extension could improve the recovery performance of the 21~cm signal
in future work.

While we demonstrated the use of RPCA and GILC method for HI signal 
recovery in mid-redshift 21~cm experiment, it could also 
be applied to other redshift range (e.g. EoR) or even other spectral line (e.g. CO)
intensity mapping experiments as well. For example, 
\cite{2016MNRAS.456.2749O} showed that the GNILC method can be used to
recover the 21~cm signal by using a prior of the theoretical HI frequency 
covariance matrix, here we have shown that actually we could extract a 
good estimate of the HI 21~cm frequency covariance matrix from 
the observed data itself by the RPCA method. This could
avoid a biased prior, which is even more useful in the EoR 21cm experiment or 
other spectral line intensity mapping experiment, since in those cases 
we have less reliable knowledge of the expected signals. 

\acknowledgments
We acknowledges the support of the MoST through grant 2016YFE0100300,
NSFC through grant No. 11473044, 1633004,  and the CAS Frontier Science Key Project No. 
QYZDJ-SSW-SLH017. 

\appendix

\section{Generate the Simulated Maps} \label{S:gen}
For the simulation of this paper, we use the software package CORA \cite{2014ApJ...781...57S,2015PhRvD..91h3514S} to generate
the input sky signals, including HI 21~cm signal and foreground
components. For the HI emission, it assumes the Planck 2013 cosmological 
model\cite{2014A&A...571A..16P}. The foreground models used in
CORA are based on \cite{2005ApJ...625..575S}. However, for
illustration purpose, here we only include two foreground
components, the Galactic synchrotron emission and extragalactic
point sources, which are the dominant components of the foreground
contaminations related to 21~cm experiments and are also relatively
accurately modeled ones. The angular power spectrum of these two
components can both be modeled in the form of
\begin{equation} \label{eq:clmdl}
  C_{l}(\nu, \nu') = A \left( \frac{l}{100} \right)^{-\alpha} \left(
    \frac{\nu \nu'}{\nu_{0}^{2}} \right)^{-\beta}
  e^{-\frac{1}{2\xi_{l}^{2}}\ln^{2}(\nu / \nu')}.
\end{equation}
We use the recalibrated model parameters for the 700-800 MHz band of HI intensity mapping 
experiment. The CORA package also implemented the polarized emission model but 
for simplicity we only consider the total intensity model. 
The parameters of the models given in \autoref{eq:clmdl}
is listed in \autoref{tab:clp}.

To generate the galactic synchrotron emission, 
CORA uses the processed 408 MHz Haslam 
map (bright point sources and striping are removed)
as an template, and extrapolate to other frequencies using a spectral 
index from the Global Sky Model (GSM) \cite{2008A&A...490.1093M},  
with a Gaussian random realization of \autoref{eq:clmdl}
that adds fluctuations in frequency and on small angular scales.  
The extragalactic point sources simulations
come from three components: a population of bright point
sources ($S > 10$~Jy at 151~MHz); a synthetic population of dimmer
sources down to 0.1~Jy at 151~MHz; and an unresolved background of
dimmer sources ($S < 0.1$~Jy) modeled as a Gaussian random
realization from Eq.~(\ref{eq:clmdl}) with the point source model
parameters listed in \autoref{tab:clp}.

On the scales of interest, the 21~cm power spectrum is given as
\begin{equation} \label{eq:PTb}
  P_{T_{b}}(\vec{k}; z, z') = \bar{T}_{b}(z) \bar{T}_{b}(z') (b + f
  \mu^{2})^{2} P_{m}(k; z, z'),
\end{equation}
where $b$ is the bias, $f$ is the growth rate, and $P_{m}(k; z, z')$
is the real-space matter power spectrum. The mean brightness
temperature, given in \cite{2008PhRvL.100i1303C}, takes the form
\begin{eqnarray} \label{eq:Tbz}
  \bar{T}_{b}(z) &=& 0.3 \times \left( \frac{\Omega_{\text{HI}}}{
      10^{-3}} \right)   \left( \frac{1 +  z}{2.5} \right)^{1/2} \nonumber\\
&&    \times  \left( \frac{\Omega_{m} + (1 + z)^{-3}
      \Omega_{\Lambda}}{0.29} \right)^{-1/2} 
     \text{mK}.
\end{eqnarray}
We adopt the typical values of CORA parameters:  
$\Omega_{\text{HI}} b = 6.2 \times 10^{-3}$  \cite{2013MNRAS.434L..46S} and $b = 1$.
The 21cm angular power spectrum is given by \cite{2007MNRAS.378..119D}
$$C_{l}(\Delta \nu) \propto \int k^{2} dk j_{l}(k r_{\nu}) j_{l}(k r_{\nu'}) P_{T_{b}}(\vec{k}; z, z'),$$
where $\Delta \nu = \nu' - \nu$. CORA calculates
a flat-sky approximation, which is accurate to the 1\% level:
\begin{equation} \label{eq:Clzz}
  C_{l}(z, z') = \frac{1}{\pi \chi \chi'} \int_{0}^{\infty}
  dk_{\parallel} \cos(k_{\parallel} \Delta \chi) P_{T_{b}}(\vec{k};
  z, z'),
\end{equation}
where $\chi$ and $\chi'$ are the comoving distances to redshift $z$
and $z'$ and $\Delta \chi$ is their difference. See Refs.
\cite{2014ApJ...781...57S,2015PhRvD..91h3514S} and  the
CORA documentation for further details.

\begin{table}[htbp]
  \centering
      \caption{Parameters for foreground model.}
  \begin{tabular}{llllll}
    \hline\hline
    Component & Polarization & $A (\text{K}^{2})$ & $\alpha$ & $\beta$ & $\xi$ \\
    \hline
    Galaxy & TT & $6.6 \times 10^{-3}$ & 2.80 &  2.8 & 4.0 \\
    Point sources & TT & $3.55 \times 10^{-4}$ & 2.10 &  1.1 & 1.0 \\
    \hline\hline
  \end{tabular}
  \label{tab:clp}
\end{table}


\bibliography{rpca}

\end{document}